\documentclass[aps,showpacs,amsmath,amssymb]{revtex4}
\usepackage{graphicx}
\usepackage{dcolumn}
\usepackage{bm}

\begin{document}



\title{Vorticity and the Nernst Effect in Cuprate Superconductors}


\author{N. P. Ong$^1$, Yayu Wang$^1$, S. Ono$^2$, Yoichi Ando$^2$, and S. Uchida$^3$}
\affiliation{
$^1$Department of Physics, Princeton University, Princeton, New Jersey 08544, USA\\
$^2$Central Research Institute of Electric Power Industry, Komae, Tokyo 201-8511, 
Japan\\
$^3$School of Frontier Sciences, University of Tokyo, Tokyo 113-8656, Japan
}

\date{\today}

\begin{abstract}
We present a review of the vortex-Nernst effect in the 3 cuprate families $\rm 
La_{2-x}Sr_xCuO_4$, $\rm Bi_2Sr_2CaCu_2O_8$, and $\rm YBa_2Cu_3O_{y}$, and discuss the 
scenario that the superconducting transition in the hole-doped cuprates corresponds to 
the destruction of long-range phase coherence rather than the vanishing of the 
order-parameter amplitude.  
\end{abstract}
\pacs{74.40.+k,72.15.Jf,74.72.-h}
\maketitle

\begin{figure*}[b]
\includegraphics[width=.35\textwidth]{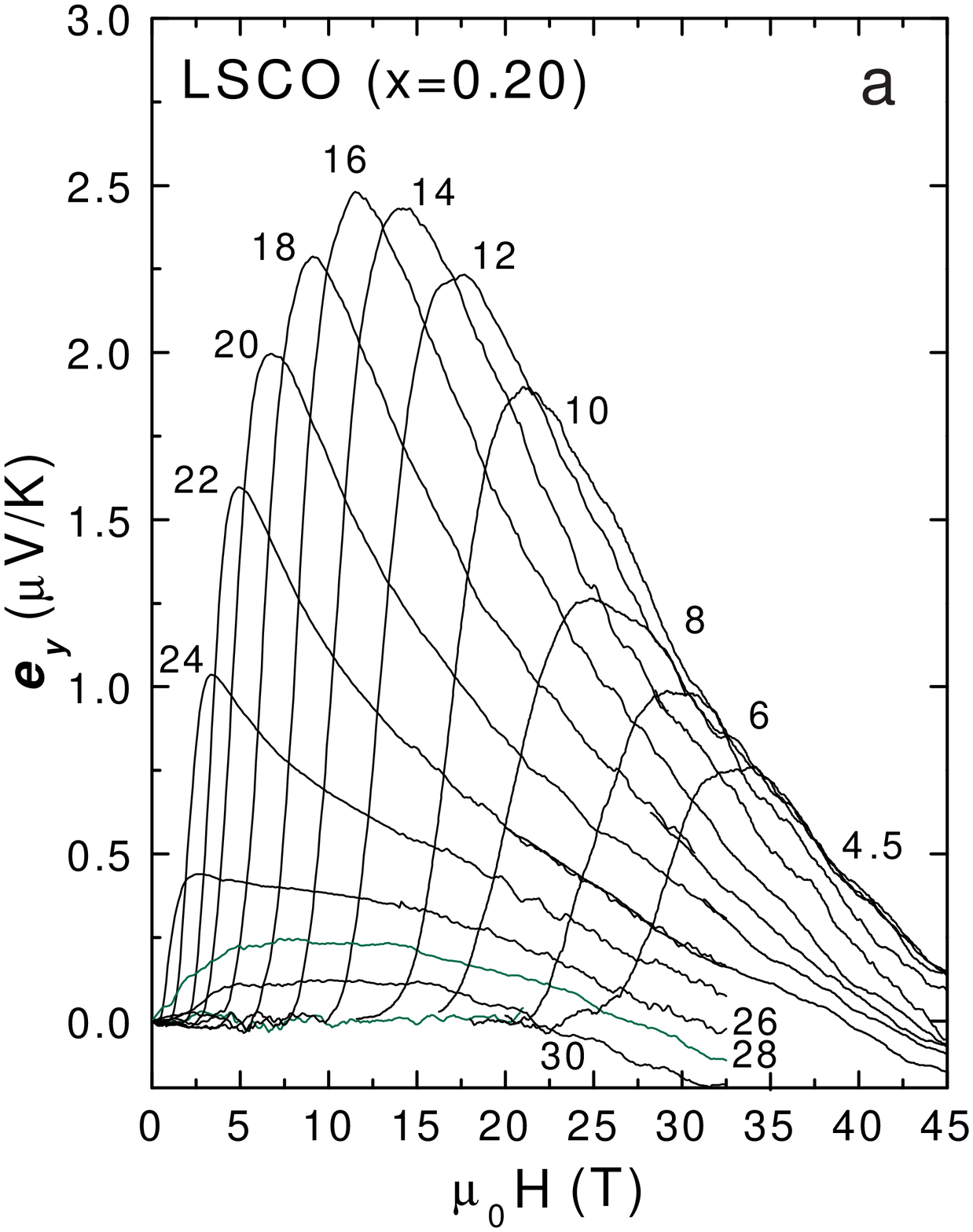}
\hfil
\includegraphics[width=.35\textwidth]{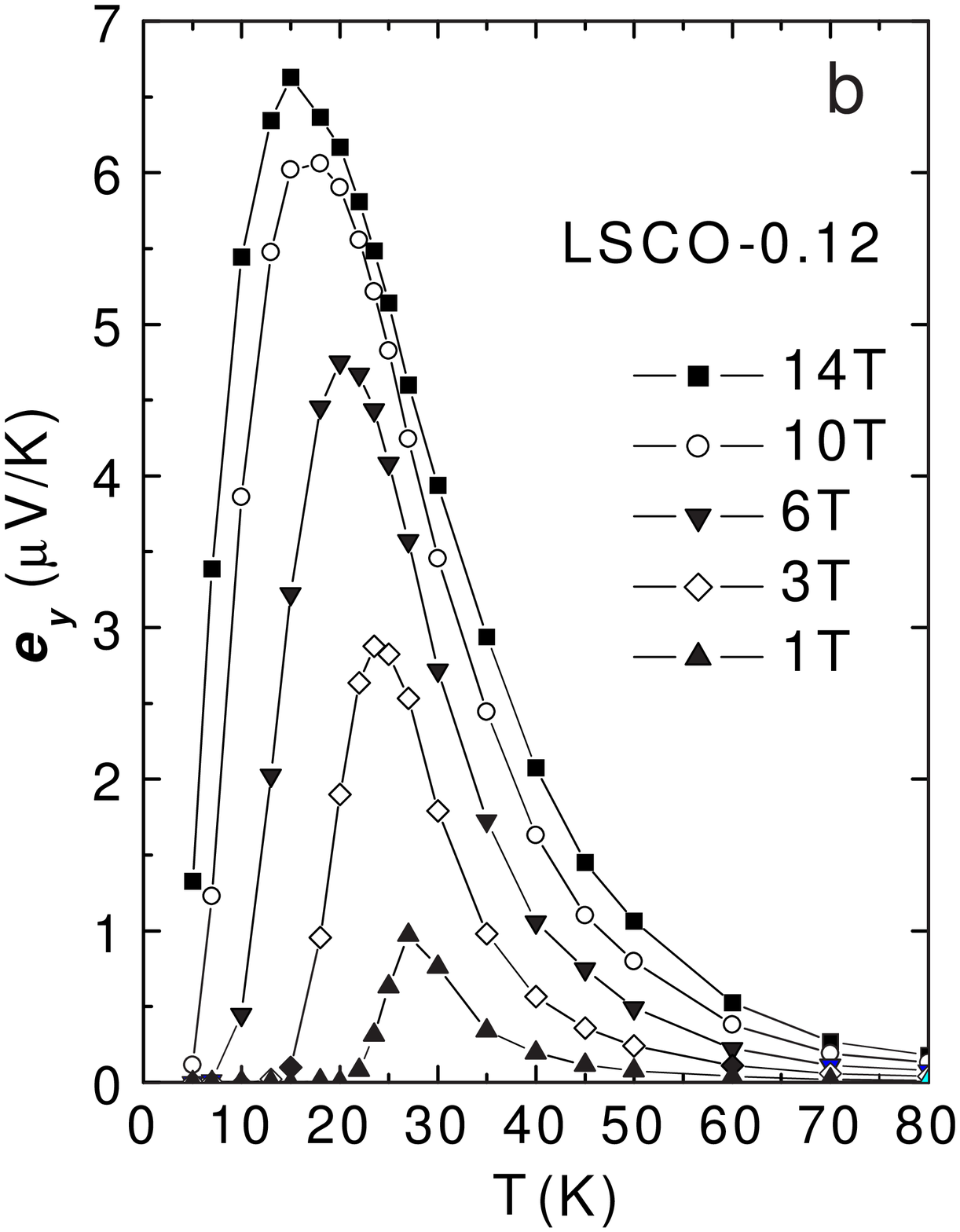}
\caption{(a) The observed Nernst signal $e_y = E_y/|\nabla T|$ vs. field $H$ up to 45 T 
in overdoped LSCO at selected $T$ ($x$ = 0.20, $T_{c0}$ = 28 K).  The prominent peak and 
decrease at high fields are the vortex Nernst signal.  At $T$ > 18 K, the Nersnt signal 
of the holes which is negative causes $e_y$ to become slightly negative at high fields.   
(b) The $T$ dependence of $e_y$ at fixed $H$ in underdoped LSCO ($x$ = 0.12).  Note the 
continuity of the signal across the transition $T_{c0}$ = 28.9 K.
}
\label{LSCONH}
\end{figure*}

In conventional superconductors, the amplitude of the complex order parameter 
$\hat{\Psi}({\bf r}) = |\Psi|\mathrm{e}^{i\theta({\bf r})}$ vanishes as the temperature 
$T\rightarrow T_{c0}^{-}$ where $T_{c0}$ is the zero-field critical temperature.  
Concurrently,  \emph{long-range} phase coherence also vanishes at $T_{c0}$.  In an 
alternate scenario, phase coherence may vanish at $T_{c0}$, but $|\Psi|$ remains finite 
to temperatures significantly higher than $T_{c0}$.  An extreme example is the 
Kosterlitz Thouless (KT) transition~\cite{KT} in 2D superconductors.  Although the 
cuprates are intrinsically 3D systems, evidence has accumulated over the years that 
their transition to superfluidity reflects the loss of phase coherence rather than the 
vanishing of $|\Psi|$.  Early $\mu$SR experiments~\cite{Uemura} showed that $T_{c0}$ 
scales linearly with the superfluid density in the underdoped regime.  In thin-film 
samples, kinetic inductance is observed to persist to $\sim$25 K above 
$T_{c0}$~\cite{Corson}.  A series of Nernst 
experiments~\cite{Xu,WangPRB,WangPRL,WangScience,OngRio,Capan,Wen} has provided perhaps 
the firmest evidence that the transition in zero field corresponds to rapid 
proliferation of mobile vortices which destroy phase coherence~\cite{Emery,Sudbo}.  We 
review these results here.

In the vortex-liquid state, a temperature gradient $-\nabla T||{\bf \hat{x}}$ induces 
vortices to flow with velocity ${\bf v|| \hat{x}}$ (the field $\bf B||\hat{z}$).  As the 
vortices cross a reference line parallel to $\bf \hat{y}$, stochastic phase slips 
generate a $dc$ voltage $V_J$ via the Josephson equation $2eV_J = 2\pi \hbar \dot{n}_v$, 
which translates to a weak electric field $E_y = Bv$ antisymmetric in $\bf B$ (here 
$\dot{n}_v$ is the average number of vortices crossing the line per second).  The Nernst 
signal, defined as $e_y = E_y/|\nabla T|$, has turned out to be a powerful probe of the 
existence of vorticity in the phase diagram of the cuprates~\cite{Xu,WangPRB}.  A recent 
theoretical discussion may be found in Ref.~\cite{Iddo}. 

\begin{figure*}[b]
\includegraphics[width=.35\textwidth]{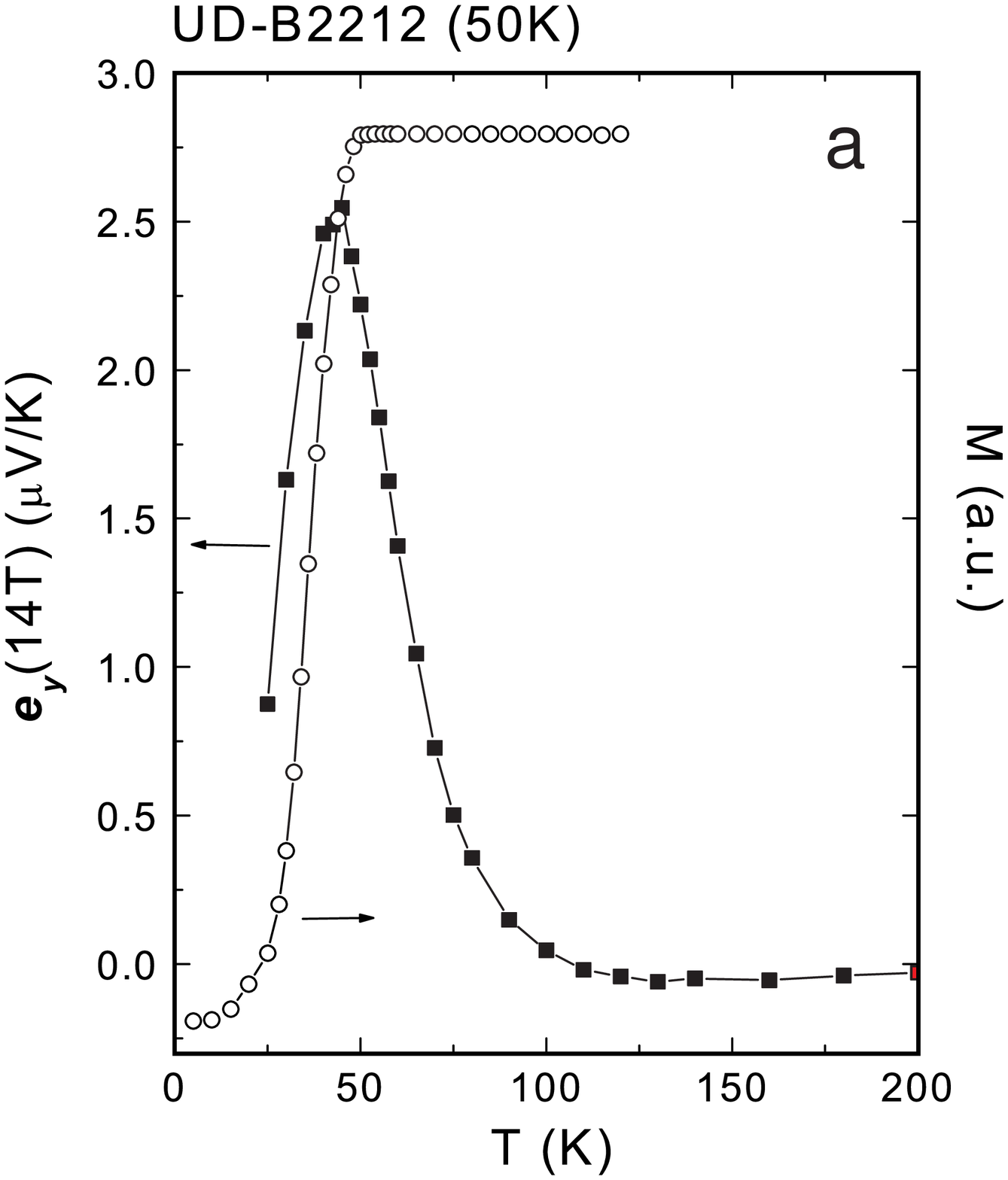}
\hfil
\includegraphics[width=.35\textwidth]{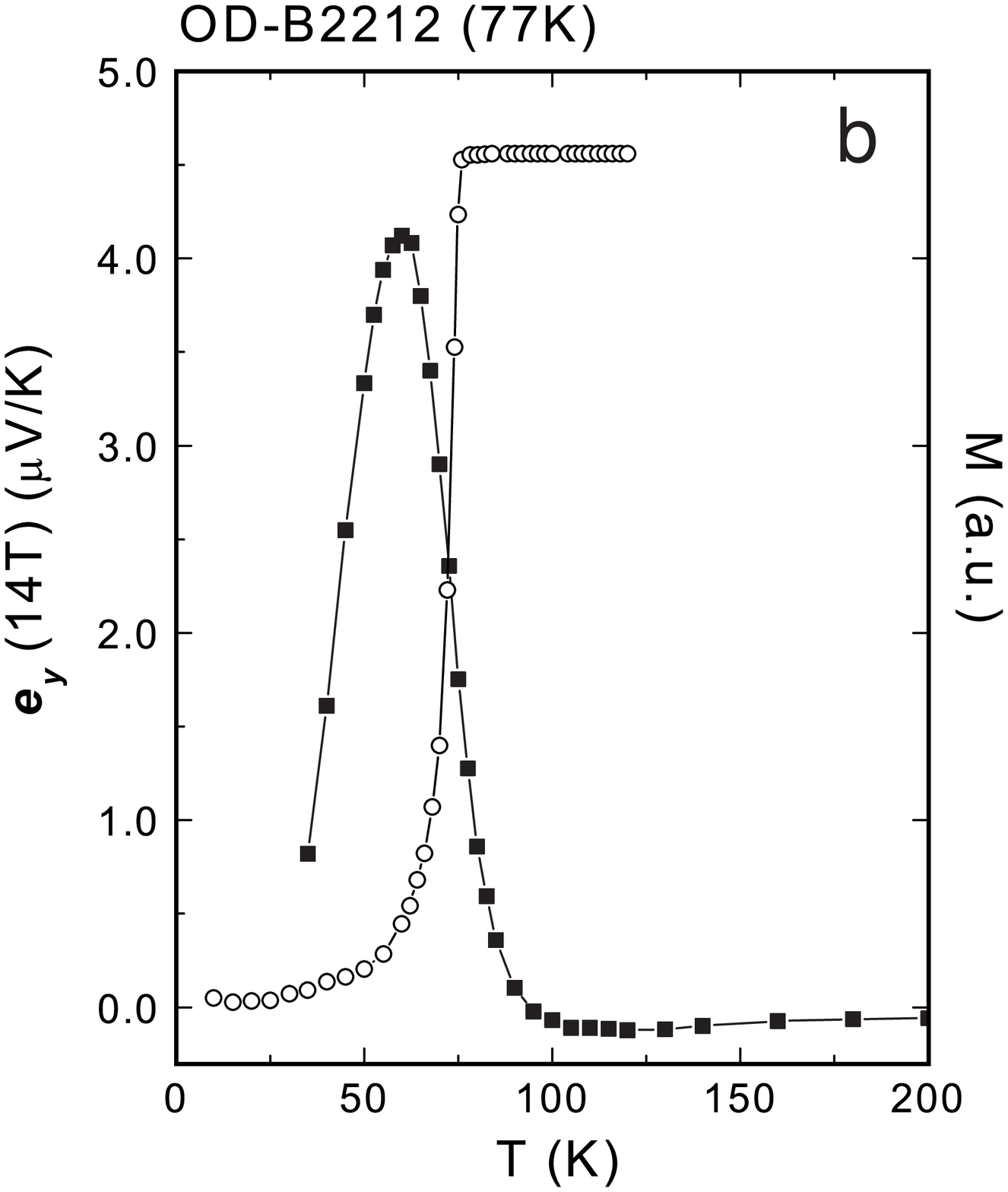}
\caption{Comparison of the Nernst signal $e_y$ measured at 14 T (solid symbols) in 
underdoped Bi 2212 (Panel a) and overdoped Bi 2212 (b).   The diamagnetic susceptibility 
$\chi$ measured with $H$ = 10 Oe (open symbols) shows the sharp Meissner transition at 
50 K and 77 K in the 2 crystals, respectively.  Above $T_{\nu}$ (onset of the vortex 
signal), the carrier contribution to $e_y$ is weak ($<$100 nV/K at 14 T) and negative.
}
\label{Bi2212}
\end{figure*}

In cuprates, $e_y(T,H)$ exists as a strong signal over a rather large region in the 
$T$-$H$ (temperature-field) plane.  Figure \ref{LSCONH}a shows plots of $e_y$ vs. $H$ in 
overdoped $\rm La_{2-x}Sr_xCuO_4$ (LSCO) in which $x = 0.20$ and $T_{c0}$ = 28 K.  The 
characteristic `tent' profile of the curve of $e_y$ vs. $H$ below $T_{c0}$ becomes 
apparent only in very high fields~\cite{WangPRL,WangScience}.  The profile is strikingly 
similar to that observed in the Abrikosov state of low-$T_c$ type II superconductors.  
Starting at the lowest $T$ (4.5 K), we see that $e_y$ is zero until the solid-liquid 
melting transition occurs at $H_m$ ($\sim 25$ T).  In the liquid state, $e_y$ rises to a 
maximum value before decreasing monotonically towards zero at a field that we identify 
with the upper critical field $H_{c2} \simeq$ 50 T (the field at which the pairing 
amplitude is completely suppressed).  As $T$ increases, both $H_{m}$ and the peak field 
$H^*$ move to lower field values.  Both the peak feature as well as the monotonic 
decrease towards zero become increasingly dramatic as $T$ is increased to the interval 
10 to 20 K.  A complication in overdoped LSCO is that the hole carriers contribute a 
moderately large, negative Nernst signal~\cite{WangPRB}.  Close to $T_{c0}$, this 
carrier contribution pulls the vortex signal to negative values in high fields.  (We 
define~\cite{WangPRB} the Nernst signal to be positive if the observed $E$-field is 
consistent with a vortex origin, i.e. given by $\bf E = B\times v$. )  The hole 
contribution complicates the task of isolating the vortex signal at high $T$ in 
overdoped samples, but is negligible for $x\le$ 0.17.   

\begin{figure*}[t]
\includegraphics[width=.35\textwidth]{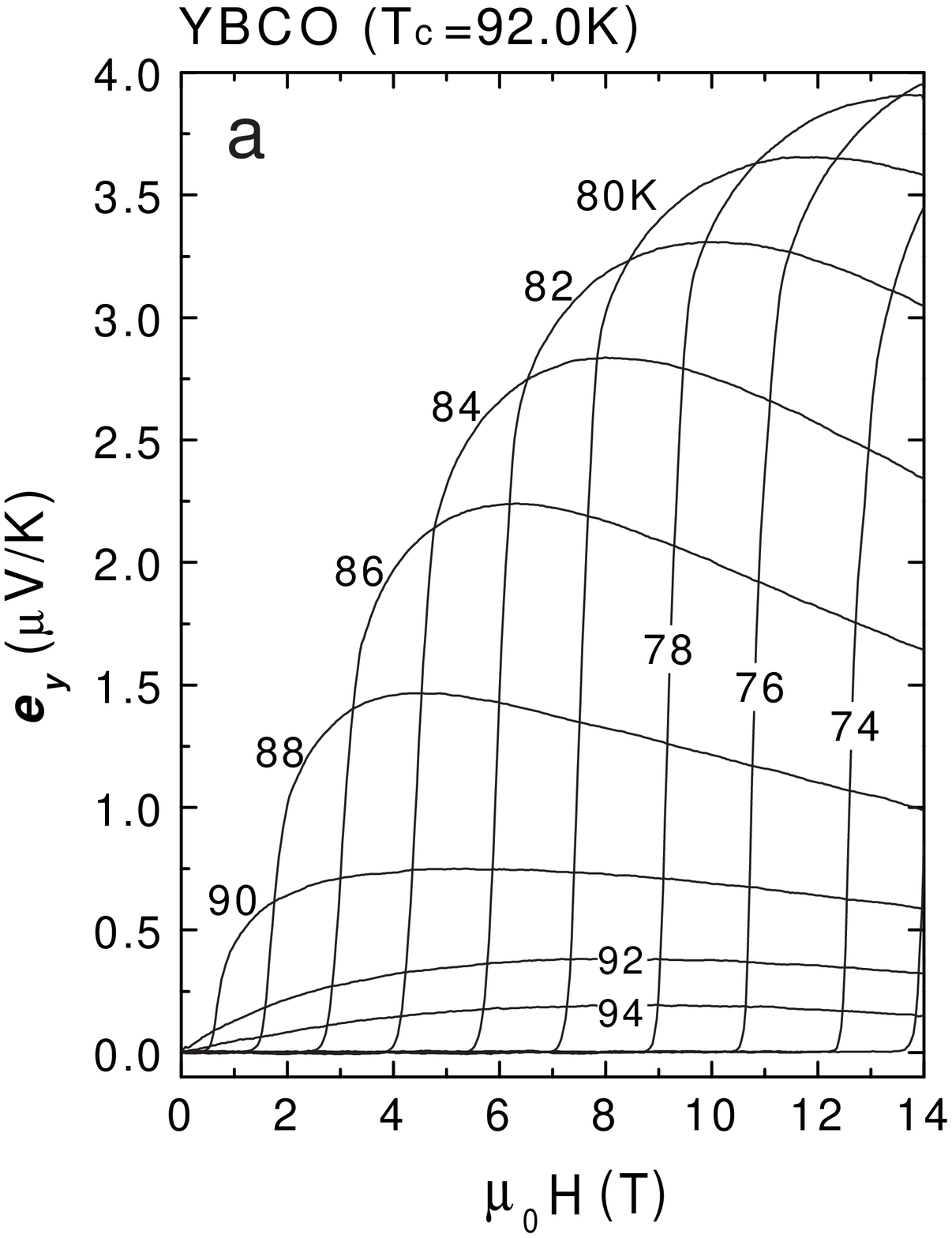}
\hfil
\includegraphics[width=.35\textwidth]{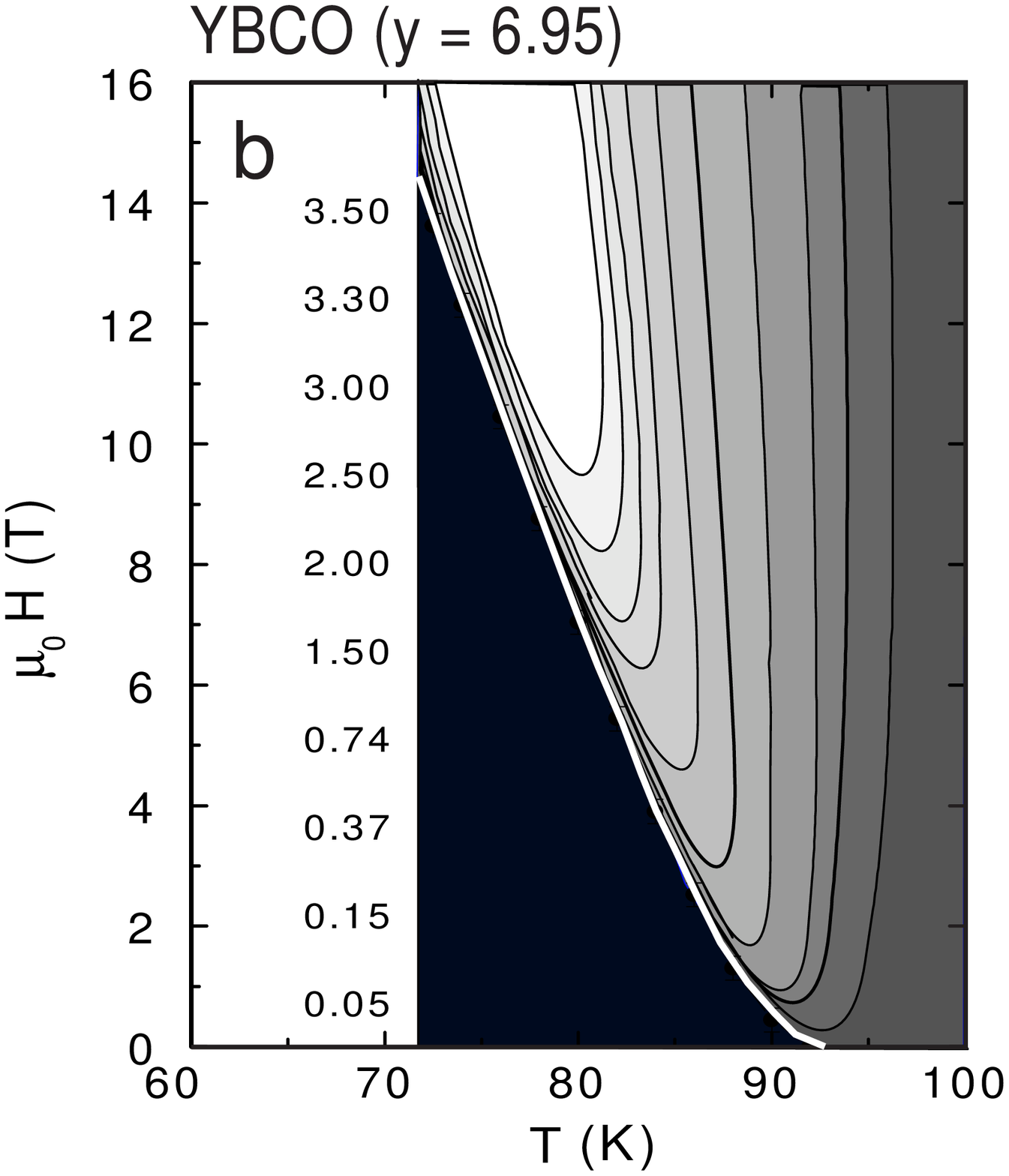}
\caption{(a) Curves of $e_y$ vs. $H$ in untwinned YBCO ($y$ = 6.99, $T_{c0}$ = 93 K) 
with $-\nabla T||{\bf b}$ fro 74 to 94 K. (b) The contour plot of $e_y(T,H)$ in the 
$T$-$H$ plane of optimally doped YBCO ($y$=6.95) with contour values indicated on the 
left column (in $\mu$V/K).  $H_m(T)$ (white curve) separates the vortex solid phase 
(shaded black) from the liquid phase.  The vortex Nernst signal onset is $T_{onset} 
\simeq$ 105 K.
}
\label{YBCO7}
\end{figure*}

A different perspective on $e_y(T,H)$ is shown in Fig. \ref{LSCONH}b in underdoped LSCO 
($x = 0.12$, $T_{c0}$ = 28.9 K).  Each curve represents the profile of $e_y$ vs. $T$ at 
fixed field.  At this doping, the hole contribution to $e_y$ is negligibly small 
compared with the vortex signal.  The important feature here is that $e_y$ extends 
continuously to $T$ high above $T_{c0}$.  There is no sign of a sharp boundary 
separating the vortex liquid state at low $T$ and $H$ from a high-$T$ `normal state'.  
Displaying the data in this way brings out clearly the smooth continuity of the vortex 
signal above and below $T_{c0}$.  This continuity is also apparent in contour plots of 
$e_y(T,H)$ in the $T$-$H$ plane, as reported in Ref. \cite{WangPRL,OngRio} (see Fig. 
\ref{YBCO7}b as well).

Of all the cuprates, the Bi-based families display the weakest carrier-Nernst signals.  
Figure \ref{Bi2212} shows the total $e_y$ measured at 14 T in underdoped $\rm 
Bi_2Sr_2CaCu_2O_8$ (Bi 2212) and overdoped Bi 2212 [Panels (a) and (b), respectively].  
In both cases, the hole signal is negative and always less 100 nV/K in magnitude (at 14 
T).  In the underdoped sample (Panel a), the vortex signal appears at 120 K and 
increases rapidly to a peak value of 2.6 $\mu$V/K near $T_{c0}$ = 50 K (the Meissner 
response is shown by the open circles).  Hence there exists a 70 K interval over which 
vortex fluctuations are readily observed.  In the overdoped sample (b), the onset is at 
100 K whereas $T_{c0}$ = 77 K.  The fluctuation regime is narrower in the overdoped 
regime (this trend is shared by LSCO), but it still extends over a 23-K interval.

\begin{figure*}[t]
\includegraphics[width=.35\textwidth]{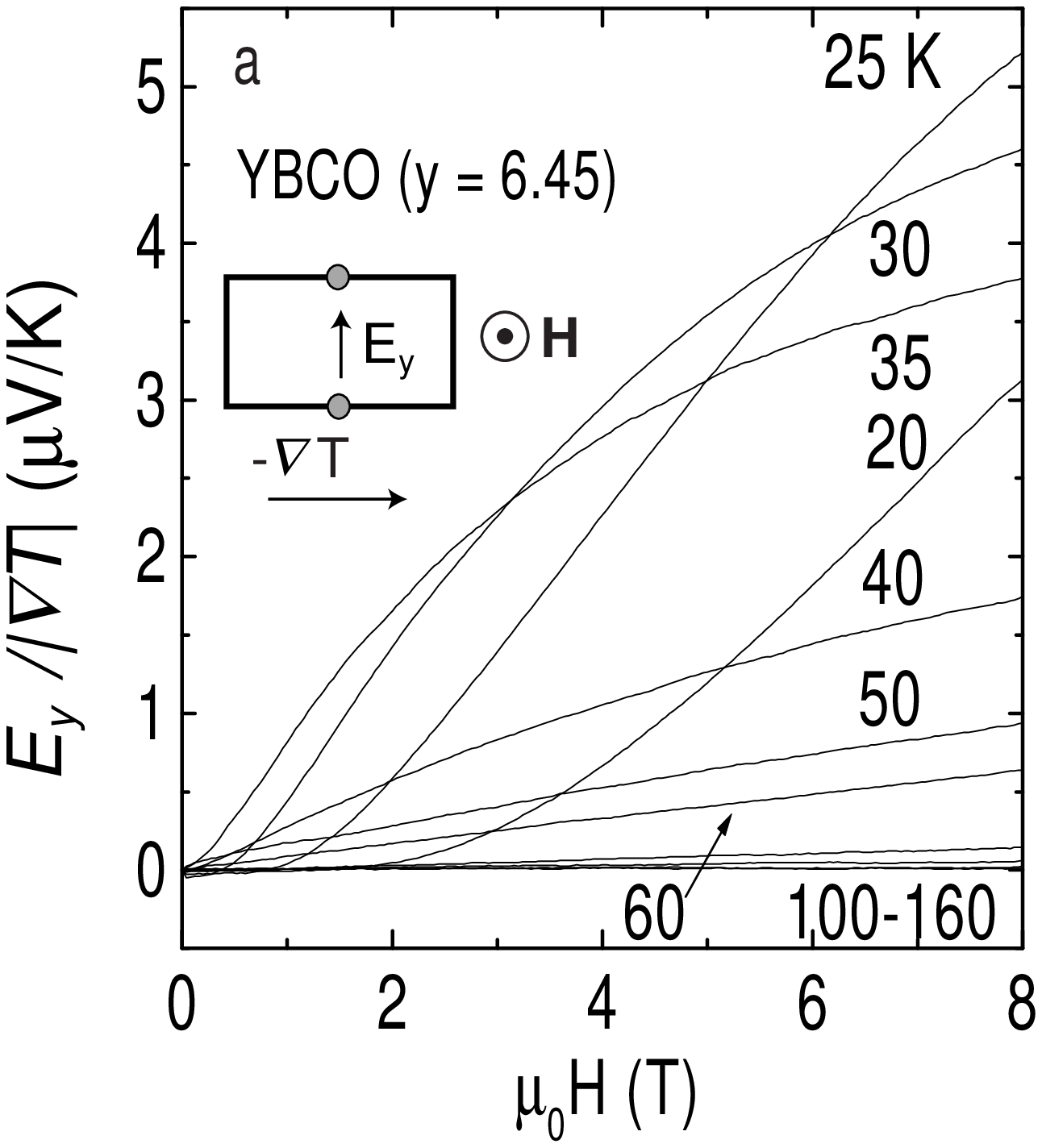}
\hfil
\includegraphics[width=.35\textwidth]{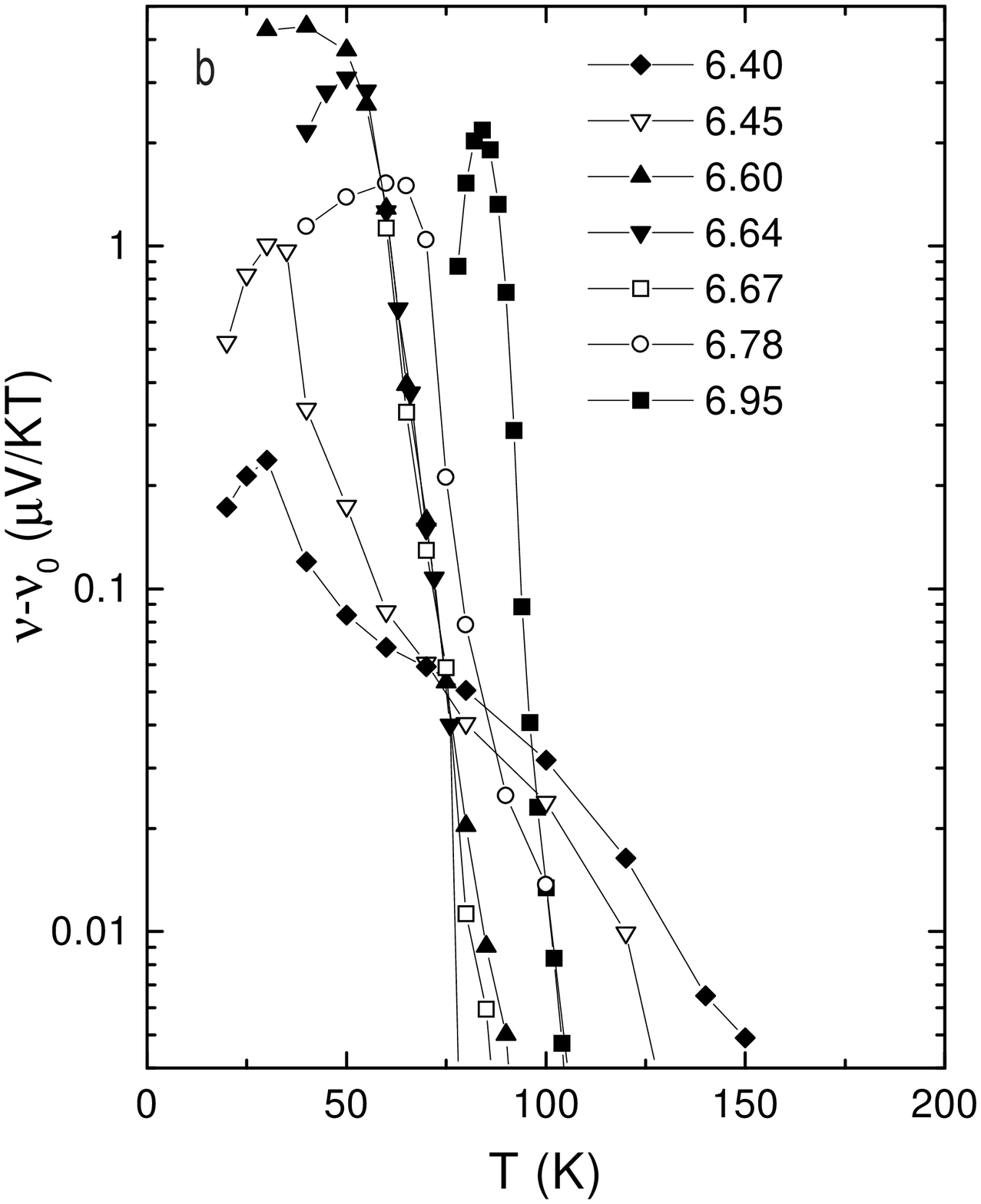}
\caption{(a) Curves of $e_y$ vs. $H$ in underdoped YBCO ($y$ = 6.45, $T_{c0}$ = 45 K) 
from 20 to 160 K.  (b) The vortex Nernst coefficient $\nu-\nu_N$ (with hole signal 
subtracted) plotted vs. $T$ in a series of YBCO crystals ($y$ = 6.40 to 6.95).  In the 
doping interval $6.60<y<6.80$, $\nu_N$ is relatively large and negative above $T_{c0}$ 
(see text).
}
\label{YBCO5}
\end{figure*}

We next discuss Nernst results in $\rm YBa_2Cu_3O_{y}$ (YBCO).   Because overdoped YBCO 
is known to have the smallest anisotropy of all the cuprates, it is interesting to 
examine its fluctuation regime.  Figure \ref{YBCO7}a displays curves of $e_y$ vs. $H$ at 
temperatures 74 to 94 K for an untwinned YBCO crystal with $y = 6.99$ and $T_{c0}$ = 93 
K with $-\nabla T||{\bf \hat{b}}$.  In contrast to LSCO and Bi 2212, the increase in 
$e_y$ at the melting line is nearly vertical.  In Fig. \ref{YBCO7}b, we display the 
variation of $e_y(T,H)$ in the $T$-$H$ plane as a contour plot (for optimally-doped YBCO 
with $y$ = 6.95).  The solid lines are contours of $e_y$ with the values given in the 
left-hand column.  The melting field $H_m$ is shown as a white curve.  To the left of 
$H_m(T)$, the black region represents the vortex-solid phase, in which $e_y = 0$.  A 
striking feature of the contour plot is that even in optimal YBCO, the fluctuation 
regime is present although it is quite narrow ($\sim$10 K).   As we decrease the oxygen 
content $y$, the fluctuation regime expands significantly, again in agreement with the 
trend in LSCO and Bi 2212.  Figure \ref{YBCO5}a displays curves of $e_y$ vs. $H$ for an 
underdoped YBCO ($y$ = 6.45) in which the vortex signal extends to above 100 K even 
though $T_{c0}$ = 45 K.  At temperatures above $T_{c0}$ where $e_y$ is nearly linear in 
$H$ for fields below 14 T, it is convenient to represent the Nernst magnitude by the 
initial slope $\nu = e_y/H$ (the Nernst coefficient).  Figure \ref{YBCO5}b displays how 
$\nu$ (with the carrier contribution $\nu_N$ subtracted) varies with $T$ in 7 crystals 
of YBCO over a broad range of $y$.  Above $T_{c0}$, the vortex Nernst coefficient 
$\nu-\nu_N$ falls monotonically with a slope that depends monotonically on the oxygen 
content.  As $y$ decreases from 6.95, the slope becomes progressively weaker until in 
the sample with $y$ = 6.40, the fluctuation regime extends to 140 K.  There exists a 
feature of the Nernst effect in YBCO that is not seen in the other hole-doped cuprates 
examined to date (LSCO, Bi 2201, Bi 2212 and Tl 2201).  In a narrow range of doping 
$6.60<y< 6.80$, the Nernst coefficient $\nu$ is strongly negative in a 10-20 K interval 
above $T_{c0}$.  This negative contribution possibly arises from holes in the chains.  
As shown in Fig. \ref{YBCO5}b, in the doping range $6.60 < y< 6.80$, the negative 
contribution causes $\nu-\nu_N$ to fall very steeply.  We believe this may be an 
artifact of an unsatisfactory subtraction procedure.  As we further decrease $y$ ($< 
6.60$), this negative contribution abruptly disappears, and we recover a pattern more 
similar to that in the other hole-doped cuprates.

\begin{figure}[h]
\includegraphics[width=.5\textwidth]{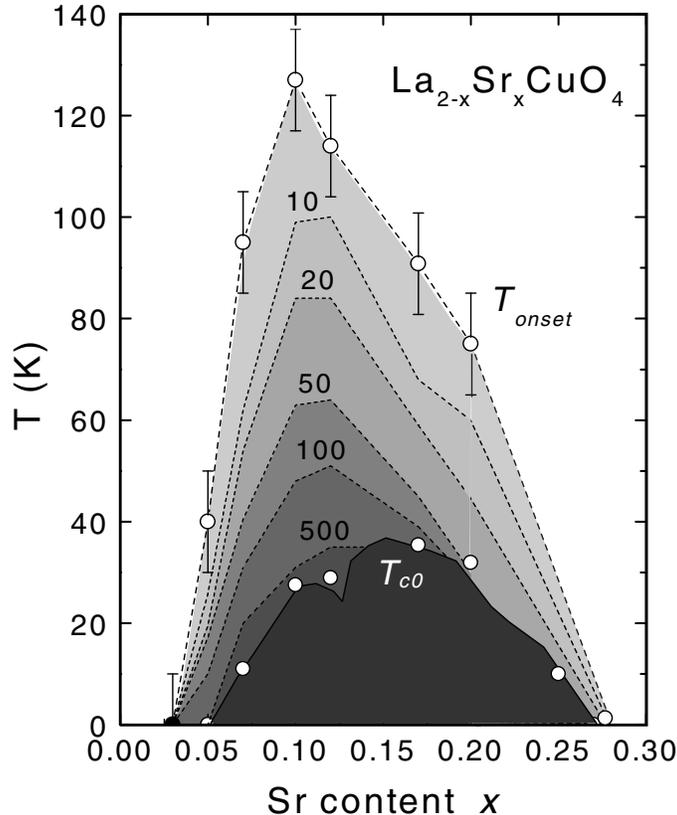}
\caption{Phase diagram of LSCO showing contour lines of the vortex Nernst signal 
observed above $T_{c0}$ ($\nu$ at each contour is given in nV/KT).  Note that the 
highest temperature occurs near $x$ = 0.10 (rather than 0.15) in all contours.  The 
vortex-Nernst signal is not observed in the samples at $x$ = 0.26 and 0.03.  
}
\label{phasedia}
\end{figure}

\emph{Discussion} The Nernst measurements, which are most extensive in LSCO, provide the 
following picture of the fluctuation regime above $T_{c0}$.  The phase diagram for LSCO 
shown in Fig. \ref{phasedia} has been pieced together from results obtained in 8 
crystals (see Refs. \cite{Xu,WangPRB,WangPRL,WangScience}).  The vortex excitation state 
(shaded) extends upwards to high temperatures above $T_{c0}$ deep into the pseudogap 
state.

We note that the vortex signal is strictly confined to the doping interval 
($0.03<x<0.26$) in which superconductivity is observed (the right anchor at $x$ = 0.26  
is a recently studied crystal in which no vortex signal is observed at any $H$ and $T$).  
This precludes other origins for $e_y$ such as density-wave excitations in the ordered 
AF state on the underdoped side or Fermi-liquid excitations on the overdoped side.  

As we cool a sample (with $x$ = 0.12, for e.g.) from high $T$, it first crosses into the 
pseudogap state at $T^*\sim$ 300 K.  Starting at 115 K, strong fluctuations appear 
between the pseudogap state and $d$-wave superconducting (dSC) state.   The existence of 
short-lived superconducting regions with short-range phase rigidity and vorticity leads 
to a weak vortex-Nernst signal.  As $T$ decreases from 115 K to $T_{c0}$ = 29 K, $e_y$ 
increases steeply (nominally as $T^{-2}$) with rapid expansion of the fluctuating 
regions.  However, at high $T$, bulk diagmagnetism is virtually undetectable, even as a 
weak fluctuation signal.  We believe this is a consequence of the weak phase stiffness 
at long length scales in this vortex fluctuation phase.  Starting about 10 K above 
$T_{c0}$, one may finally observe a rapidly divergent bulk diagmagnetism signal as the 
Meissner transition $T_{c0}$ is approached and long-range phase coherence spreads 
throughout the sample (provided $H$ is zero or very weak).  A characteristic of the 
cuprates is that the vortex Nernst signal is observed high above the temperature at 
which conventional fluctuating diamagnetism can be detected.  As apparent in Fig. 
\ref{LSCONH}b, at moderately strong field, $e_y$ continues to increase smoothly through 
$T_{c0}$, attaining a maximum at lower $T$.  Below the peak temperature, $e_y$ decreases 
steeply as the vortices become less and less mobile close to the melting line.  

Broadly speaking, there seem to be 3 viewpoints regarding the nature of the pseudogap 
state:  1) It is simply a strongly fluctuating superconducting state with complete 
absence of long-range phase coherence.  2) It has an entirely different electronic state 
from dSC and competes with it for FS area.  3) It is a pairing state that is distinct 
from dSC, yet closely related to it in terms of symmetry.  Our experiments favor the 
third scenario.  The onset of the pseudogap state at $T^*$ is invariably at a much 
higher $T$ (by a factor of $\sim 2$) than the Nernst signal onset $T_{onset}$.  
Moreover, the field scale for suppressing vorticity $H_{c2}$, while high, is still 
significantly lower than field scales associated with suppresion of the pseudogap 
measured by Shibauchi \emph{et al.}~\cite{Shibauchi}.  Hence the pseudogap state seems 
to be distinct in electronic character from dSC.  Nonetheless, the vortex fluctuation 
phase displayed in Fig. \ref{phasedia} implies that fluctuations between them extend 
over a very broad range of $T$.  The strong fluctuations suggest that the two states are 
closely related in free energy.  

Finally, we return to the nature of the transition at $T_{c0}$.  If we plot $H_m$ vs. 
$T$ in the $T$-$H$ plane we find that the melting temperature $T_m(H)$ terminates at 
$T_{c0}$, i.e. $T_m(0) = T_{c0}$.  In all cuprates, the high-temperature limit of the 
vortex-solid coincides with the Meissner transition (see for e.g., Fig. \ref{YBCO7}b).  
The Nernst picture provides a fresh perspective on this feature.  Let us decrease $T$ 
from high above the melting line with $H$ fixed at 1 Tesla.  Initially, the rapidly 
mobility of the vortices (both spontaneous and field-induced) preclude long-range phase 
coherence from developing.  As soon as the melting line is crossed, the vortices become 
immobile in the solid phase and long-range phase coherence extends throughout the 
sample.  Now we repeat the cool-down at successively smaller $H$.  Because of the 
seamless continuity of the vortex-liquid, the same situation obtains in weak fields.  In 
the limit $H = 0$, we initially have equal spontaneous populations of up and down mobile 
vortices.  As we cross the terminal point $T_m(0)$, they become immobile in the vortex 
solid phase.  Hence the occurence of long-range phase coherence is also associated with 
the solid-liquid transition just as in finite fields.  In all the hole-doped cuprates, 
we seem to have the situation $T_m(0) > T_{KT}$, where $T_{KT}$ is the intrinsic KT 
transition transition in the isolated CuO$_2$ plane with the same hole concentration.   
In bulk crystals, the 3D melting transition pre-empts the KT transition~\cite{OngRio}.  
The Meissner transition is the point at which the mobile vortices become frozen in the 
solid phase, rather than a true 2D KT transition.

High-field measurements were performed at the U.S. National High Magnetic Field Lab., 
Tallahassee, a facility supported by the U.S. National Science Foundation (NSF) and the 
State of Florida.  The research at Princeton is supported by the NSF (Grant DMR DMR 
0213706).  S. U. and N.P.O. acknowledge partial support from New Energy and Industrial 
Technol. Development Org. NEDO (Japan).

\end{document}